\documentclass[twocolumn,ammsymb,amsmath,amssymb,prb,superscriptaddress,floats,showkeys,showpacs,a4paper]{revtex4-1}

\usepackage[T1]{fontenc}
\usepackage{bm}
\usepackage{graphicx}
\usepackage{natbib}
\usepackage[euler]{textgreek}
\usepackage{ulem}
\usepackage{float}
\usepackage{color}
\usepackage{verbatim}

\begin{document}

\title{Raman scattering in few--layer MoTe$_{2}$}

\author{M. Grzeszczyk}
\email{Magdalena.Grzeszczyk@fuw.edu.pl}
\affiliation{Faculty of Physics, University of Warsaw, ul. Pasteura 5, 02-093 Warszawa, Poland}

\author{K. Go\l{}asa}
\affiliation{Faculty of Physics, University of Warsaw, ul. Pasteura 5, 02-093 Warszawa, Poland}

\author{M. Zinkiewicz}
\affiliation{Faculty of Physics, University of Warsaw, ul. Pasteura 5, 02-093 Warszawa, Poland}

\author{K. Nogajewski}
\affiliation{Laboratoire National des Champs Magn\'etiques Intenses, CNRS-UJF-UPS-INSA, 25, avenue des Martyrs, 38042 Grenoble, France} 

\author{M. R. Molas}
\affiliation{Laboratoire National des Champs Magn\'etiques Intenses, CNRS-UJF-UPS-INSA, 25, avenue des Martyrs, 38042 Grenoble, France} 

\author{M. Potemski}
\affiliation{Laboratoire National des Champs Magn\'etiques Intenses, CNRS-UJF-UPS-INSA, 25, avenue des Martyrs, 38042 Grenoble, France} 

\author{A. Wysmo\l{}ek}
\affiliation{Faculty of Physics, University of Warsaw, ul. Pasteura 5, 02-093 Warszawa, Poland}

\author{A. Babi\'nski}
\affiliation{Faculty of Physics, University of Warsaw, ul. Pasteura 5, 02-093 Warszawa, Poland}

\date{\today}

\begin{abstract}

We report on room-temperature Raman scattering measurements in few--layer crystals of exfoliated molybdenum ditelluride (MoTe$_{2}$) performed with the use of 632.8 nm (1.96 eV) laser light excitation. 
In agreement with a recent study reported by G. Froehlicher et al\cite{froehlicher} we observe a complex structure of the out-of-plane vibrational modes (A$_{1\text{g}}$/A$^{'}_{1}$), which can be explained in terms of interlayer interactions between single atomic planes of MoTe$_{2}$.  
In the case of low-energy shear and breathing modes of rigid interlayer vibrations, it is shown that their energy evolution with the number of layers can be well reproduced within a linear chain model with only the nearest neighbor interaction taken into account. 
Based on this model the corresponding in-plane and out-of-plane force constants are determined.  
We also show that the Raman scattering in MoTe$_{2}$ measured using 514.5 nm (2.41 eV) laser light excitation results in much simpler spectra.  
We argue that the rich structure of the out-of-plane vibrational modes observed in Raman scattering spectra excited with the use of 632.8 nm laser light results from its resonance with the electronic transition at the M point of the MoTe$_{2}$ first Brillouin zone.

\end{abstract}

\maketitle


\section{Introduction \label{sec:Intro}}

Thin layers of transition metal dichalcogenides (TMDs) MX$_{2}$, where M = transition metal, X = S, Se, Te, are gaining much interest due to their unique physical and optical properties.\cite{butler}
Their crystal structure is characterized by strong ion-covalent bonds within planes of hexagonally arranged X and M atoms and by weak out-of-plane van der Waals interactions between the planes. 
The properties of the semiconducting TMDs strongly depend on the number of layers which form the structure. 
In particular the transition from the indirect bandgap in bulk TMDs to the direct bandgap in their two-dimensional form \cite{kuc} makes them promising candidates for several optoelectronic applications.\cite{jariwala} 
Important for those applications is also a wide range of the energy bandgaps, which vary with the choice of a chalcogen atom in the semiconducting TMD compound. Monolayer (1L) molybdenum ditelluride (MoTe$_{2}$) with a direct bandgap of approx. 1.1 eV \cite{ruppert,lezama} conveniently complements monolayers of other dichalcogenides of higher energy bandgaps such as MoS$_{2}$ \cite{splendiani}, MoSe$_{2}$ \cite{tongay} or WS$_{2}$ \cite{zhao} rewarding the efforts to cover with them an energy spectrum as broad as possible.
This justifies the investigation of the basic properties of few-layer MoTe$_{2}$ and in particular studies of its lattice dynamics. 
As it is well known phonons are important for several physical processes like e.g. carrier scattering \cite{kaasbjerg}, heat propagation \cite{liu} and mechanical strength of crystals. \cite{li}
For studying them in layered materials Raman scattering spectroscopy is a common technique of choice \cite{zhang}, which results from its sensitivity to strain \cite{rice} or heating \cite{sahoo}. 
The Raman scattering spectra of few-layer TMDs have been shown to strongly depend on the sample thickness. 
This property makes this method a valuable complement to atomic force microscopy in determining the number of layers constituting thin flakes of TMD crystals under study. \cite{lee,berkdemir, zhao2}
Important information on the electron-phonon interactions can also be provided by resonant Raman spectroscopy \cite{livneh, carvalho, lee2, golasa} in which the resonances between excitation or scattered photons and electronic excitations is exploited. 
Motivated by a recent observation by Guo et al. \cite{guo} that resonance of this type may be anticipated in a monolayer MoTe$_{2}$ at the energy of 2.07~eV, we have investigated the Raman scattering in few-layer MoTe$_{2}$ using He-Ne laser light with a wavelength of $\lambda$=632.8 nm for excitation. 
The energy of such excitation light, equal to 1.96~eV,  matches the energy of the vertical electronic excitation from the first valence band to the second lowest conduction band at the M point of the MoTe$_{2}$ first Brillouin zone. 
Our results obtained with the 632.8 nm light excitation are in agreement with the recently reported studies of G. Froehlicher et al. Ref. \citenum{froehlicher}. 
To highlight the resonant effect of the 632.8~nm light excitation we also measured the Raman spectra excited with 514.5 nm laser light (2.41 eV). 
We show that the spectra excited with the high energy light, resulting from out-of-plane vibrations, are much simpler than those excited using longer wavelength.
The analysis of the spectra due to the Stokes and the anti-Stokes modes of the Raman scattering confirms the resonant character of the 632.8 nm light excitation.

The $\alpha$-MoTe$_{2}$ crystal investigated in this work has the trigonal prism structure similarly to other group VI B dichalcogenides such as MoS$_{2}$, MoSe$_{2}$, WS$_{2}$ or WSe$_{2}$. 
In the bulk form MoTe$_{2}$ belongs to D$^{4}_{6h}$ space group. \cite{verble}
The symmetries of the lattice vibrations at the $\Gamma$ point of the Brillouin zone can be assigned to the following 12 irreducible representations:

\begin{equation*}
\Gamma \equiv A_{1\text{g}} \oplus 2 A_{2\text{u}} \oplus 2 B_{2\text{g}} \oplus E_{1\text{g}} \oplus 2 E_{1\text{u}} \oplus E_{2\text{u}} \oplus B_{1\text{u}} \oplus 2 E_{2\text{g}}
\end{equation*}

The acoustical modes are of A$_{2\text{u}}$ and E$_{1\text{u}}$ symmetry. 
The infrared-active modes for the in-plane (E$_{\parallel}$) and out-of-plane (E$_{\perp}$) electric field are of A$_{2\text{u}}$ and E$_{1\text{u}}$ symmetry, respectively.

There are four Raman-active modes: A$_{1\text{g}}$, E$_{1\text{g}}$, and two E$_{2\text{g}}$. 
Four other modes: B$_{1\text{u}}$, two B$_{2\text{g}}$, and E$_{2\text{u}}$ are optically inactive.

\begin{figure}[h]
\includegraphics[width=.47\textwidth]{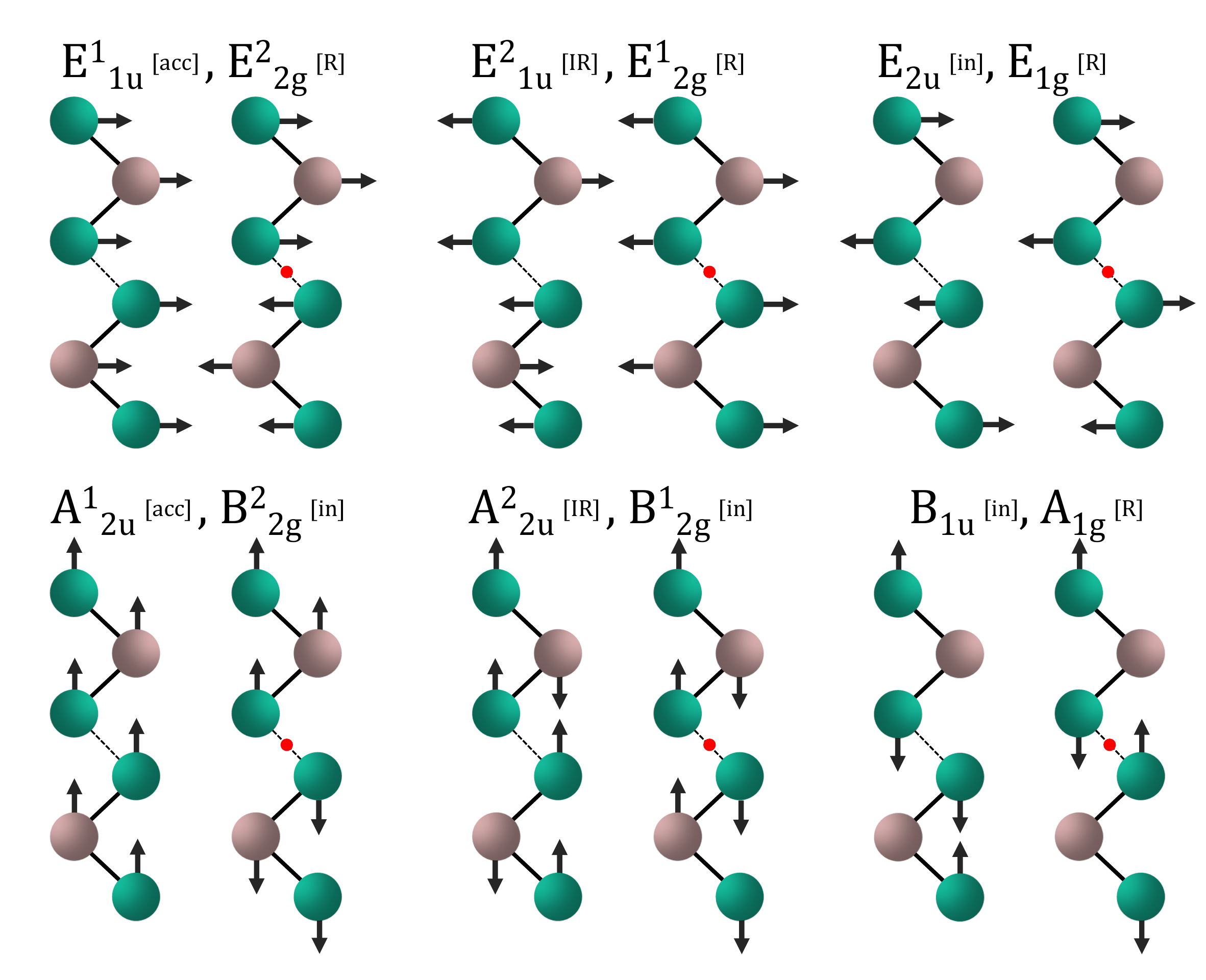}

\caption{(color online) Schematic representation of vibrational modes in MoTe$_{2}$. Conjugate pairs of modes are grouped together. 
The red points mark the inversion centers.}
\label{fig:mody}
\end{figure}

All the vibrations can be grouped into six pairs of conjugate modes. 
One of the modes in each pair is symmetric with respect to the inversion center between the nearest layers of the crystal and the other mode is antisymmetric (for the schematic representation of the vibrational modes see Fig. \ref{fig:mody}). 
The crystal symmetry along the $c$-axis is reduced in thin layers of MX$_{2}$ compounds. 
The optical activity of particular vibrational modes of these systems is also  modified as a result of the $N$-parity changes, where $N$ indicates the number of layers. 
An example of such evolution can be observed for the B$^{1}_{2\text{g}}$ mode in MoTe$_{2}$, which is inactive in bulk crystals, becomes Raman-active in few-layer samples and eventually transforms into the IR-active mode A$^{''}_{2}$ in the MoTe$_{2}$ monolayer. \cite{yamamoto}

\section{Experimental details \label{sec:procedure}}

The investigated flakes were placed on a Si/(100 nm) SiO$_{2}$ substrate. The thin MoTe$_{2}$ layers were prepared by the polydimethylosiloxan-based exfoliation technique from bulk MoTe$_{2}$ crystals purchased from HQ Graphene. 
The thicknesses of the flakes were determined from their optical contrast, previously calibrated on other MoTe$_{2}$ samples with the aid of atomic force microscopy. 
Optical spectroscopic measurements were performed in the backscattering geometry using 632.8 nm (1.96 eV) excitation with a He-Ne laser. 
The 514.5 nm (2.41 eV) line of an Ar$^{+}$ ion laser was also used for comparative studies. 
The laser light power on the sample was $\sim$100~$\mu$W and the laser spot on the sample was $\sim$1~$\mu$m in diameter.
A long working distance (x50) objective was utilized to both excite the sample and to collect the emitted light. 
The collected spectra were dispersed by a 0.75~m spectrometer equipped with 1500 grooves/mm and 2000~grooves/mm gratings and detected with a multichannel high-resolution Si charge-coupled device.


\section{Results and discussion \label{sec:pl}}

\begin{figure}[h]
\includegraphics[width=.49\textwidth]{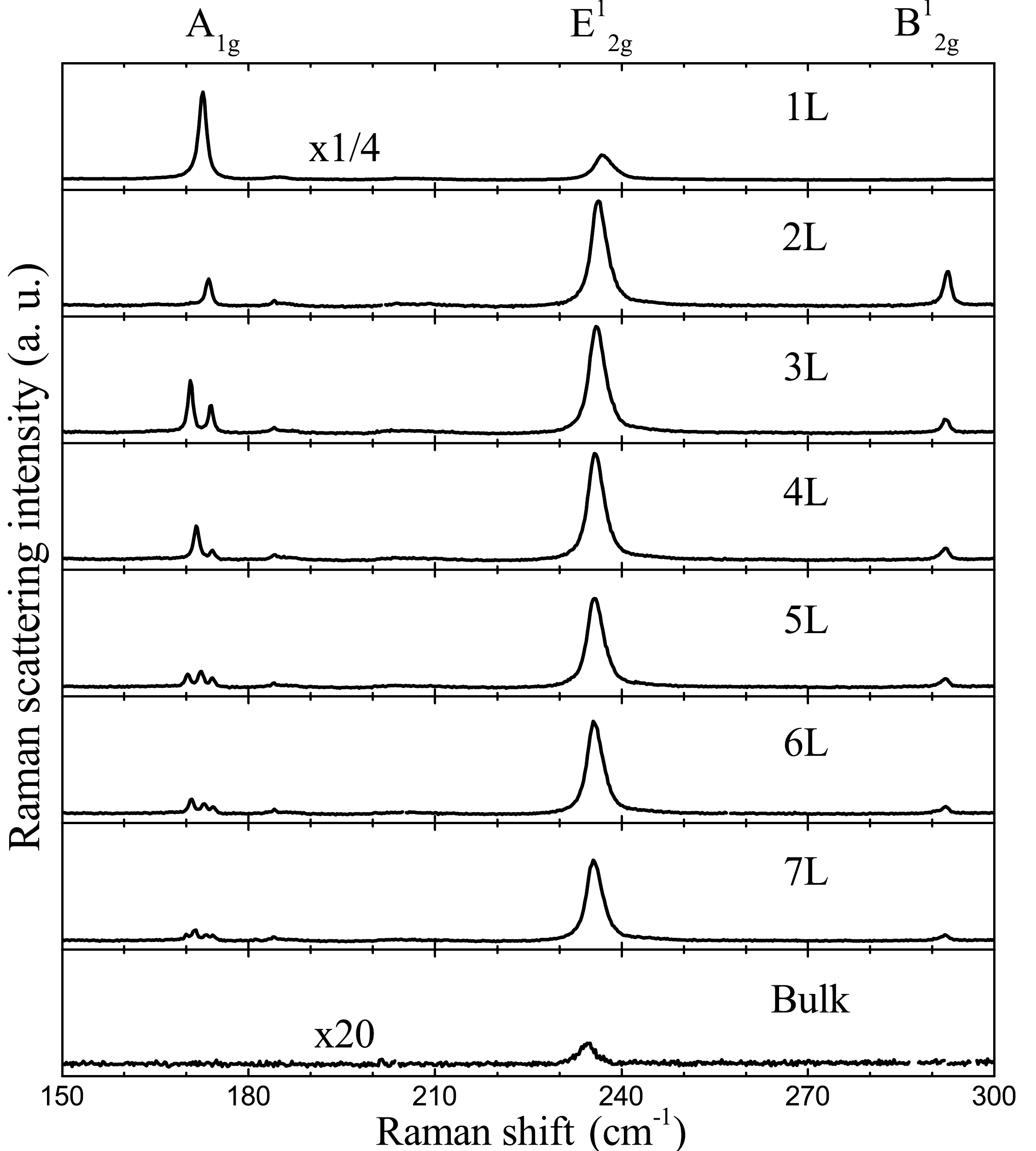}

\caption{Room-temperature Raman scattering spectra measured for different flake thickness, from mono- up to septa-layer and bulk MoTe$_{2}$ in the energy range between 150 cm$^{-1}$ and 300 cm$^{-1}$ using the 1.96 eV (632.8 nm) excitation energy.}

\label{fig:widmo}
\end{figure}

Let us focus first on the Raman scattering spectra in the energy range extending from $\sim$150 cm$^{-1}$ to $\sim$300~cm$^{-1}$. 
There are three major peaks observed in that range, which are due to first-order transitions at the $\Gamma$ point of the Brillouin zone: the out-of-plane mode A$_{1\text{g}}$ at $\sim$170 cm$^{-1}$, the in-plane mode E$^{1}_{2\text{g}}$ at $\sim$234 cm$^{-1}$ \vspace{0.5mm} \cite{yamamoto, wieting, sugai, agnihotri} and the bulk-inactive mode B$^{1}_{2\text{g}}$ at $\sim$290 cm$^{-1}$ \vspace{1mm} \cite{yamamoto} (see Fig. \ref{fig:widmo}). 
The main E$^{1}_{2\text{g}}$ peak in bulk MoTe$_{2}$ (35 nm thick in our case) corresponds to the in-plane vibrations of two Te atoms with respect to the Mo atom (see Fig. \ref{fig:mody}). 
As can be seen, the energy of this mode systematically increases with decreasing layer's thickness. 
Such behavior has already been observed in MoTe$_{2}$. \cite{froehlicher, yamamoto}
It is also characteristic for other TMDs like e.g. MoS$_{2}$ \cite{lee} or MoSe$_{2}$ \cite{tonndorf}.

The B$^{1}_{2\text{g}}$ vibrational mode is optically inactive in bulk MoTe$_{2}$. \cite{yamamoto, wieting, sugai, agnihotri}
In that mode, both Te atoms in each layer move at particular time in the same direction while the central Mo atom moves in the opposite direction (see Fig. \ref{fig:mody}).
The Mo atoms in adjacent layers vibrate out-of-phase in the mode. 
The conjugate couple to this mode is IR-active (A$^{2}_{2\text{u}}$) with Mo atoms in adjacent layers vibrating in-phase with each other. 
None of the modes can be observed in the Raman spectrum of bulk MoTe$_{2}$. 
A single vibrational mode (A$^{''}_{2}$), which corresponds to the B$^{1}_{2\text{g}}$ mode in bulk, is expected in the monolayer. \cite{yamamoto} The A$^{''}_{2}$ mode is IR-active and no peak related to this mode occurs in the Raman scattering spectrum of monolayer MoTe$_{2}$ at room temperature. 
As predicted by the group theory and previously observed \cite{yamamoto}, the mode corresponding to B$^{1}_{2\text{g}}$ in bulk becomes Raman active in few-layer MoTe$_{2}$ in the spectra for $N \geq 2$ at $\sim$290 cm$^{-1}$ (see Fig. \ref{fig:widmo}).

\begin{figure}[h]
\includegraphics[width=.5\textwidth]{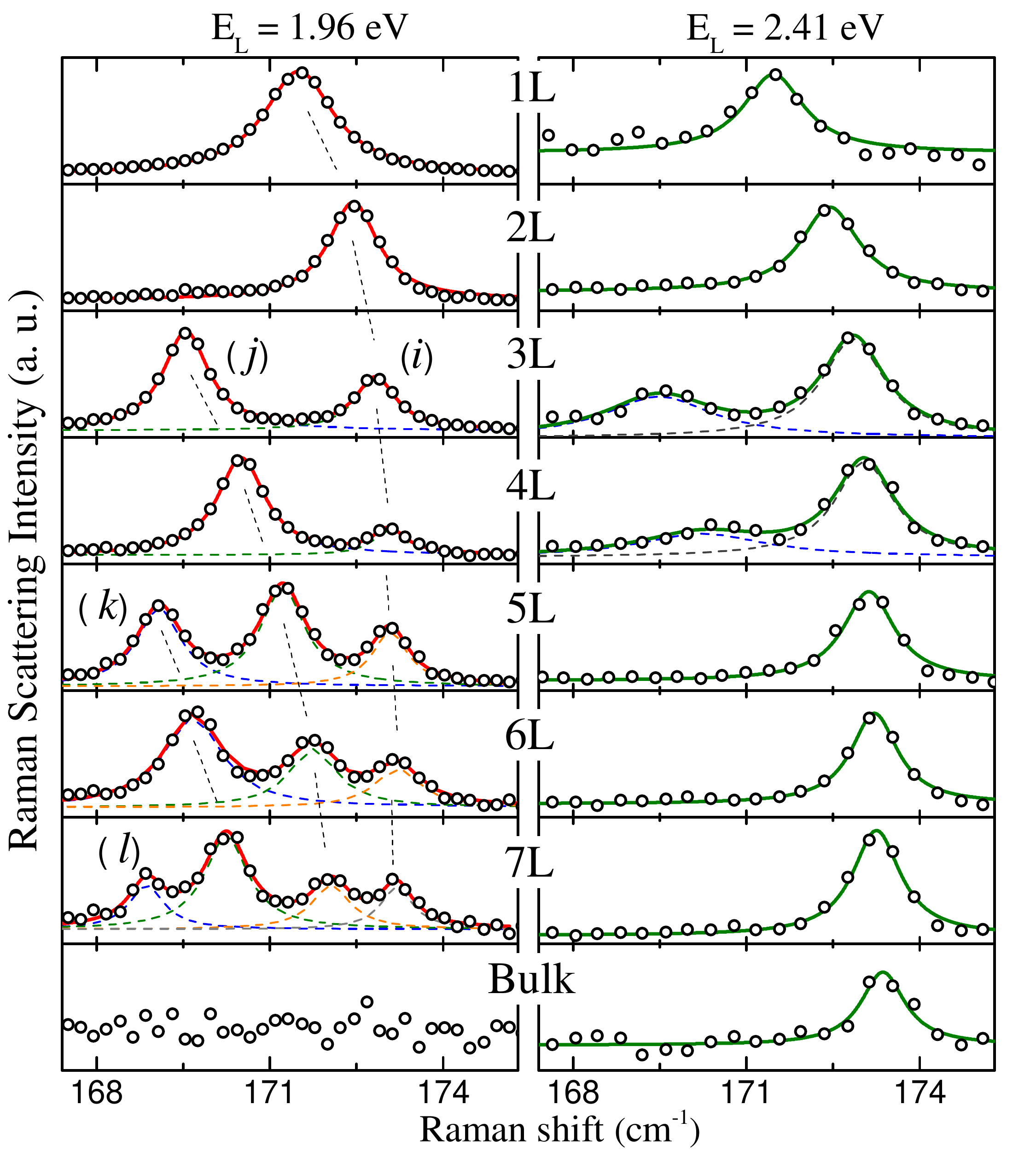}

\caption{(color-online) Features related to the out-of-plane modes in the Raman scattering spectra of monolayer (1L) to septalayer (7L) and bulk MoTe$_{2}$ measured at room temperature with the use of 632.8 nm (1.96 eV) and 514.5 nm (2.41~eV) laser light excitation. The solid red and green curves represent the results of fitting the experimental data with a superposition of Lorentzian curves, each one drawn with dashed lines.}

\label{fig:porownanie}
\end{figure}

More complex is the effect of the sample thickness on the out-of-plane modes. In few-layer MoTe$_{2}$ the number and the lineshape of spectral features, which correspond to the A$_{1\text{g}}$ mode in bulk, strongly depend on the number of layers (see Fig. \ref{fig:porownanie}). 
A general tendency, in the case of 632.8 nm excitation, is that their number increases with the flake's thickness. Such evolution can be understood in terms of interactions between the nearest-neighbor layers of a TMD crystals, as it was already observed in the case of MoSe$_{2}$ \cite{tonndorf} and recently reported for MoTe$_{2}$ (Ref. \citenum{froehlicher}). In what follows we show that the relative phase of the out-of-plane vibrations of atoms belonging to the adjacent layers of a given TMD flake is crucial for this effect. 

In bulk MoTe$_{2}$ there are two vibrational modes that correspond to the out-of-plane oscillations of Te atoms in opposite directions with respect to the central Mo atom. 
There is no phase difference between the vibrations of Te atoms in adjacent planes in the A$_{1\text{g}}$ Raman-active mode. 
This means that at particular time the Te atoms move away from the center Mo atom in all of the planes. 
A 180$^{\text{o}}$ phase difference between the vibrations of Te atoms in adjacent planes characterizes the inactive mode B$_{1\text{u}}$, where the Te atoms in a particular plane move away from the center Mo atom while the Te atoms in the adjacent planes move towards the corresponding central Mo atoms and vice versa.

\begin{figure}[h]
\includegraphics[width=.39\textwidth]{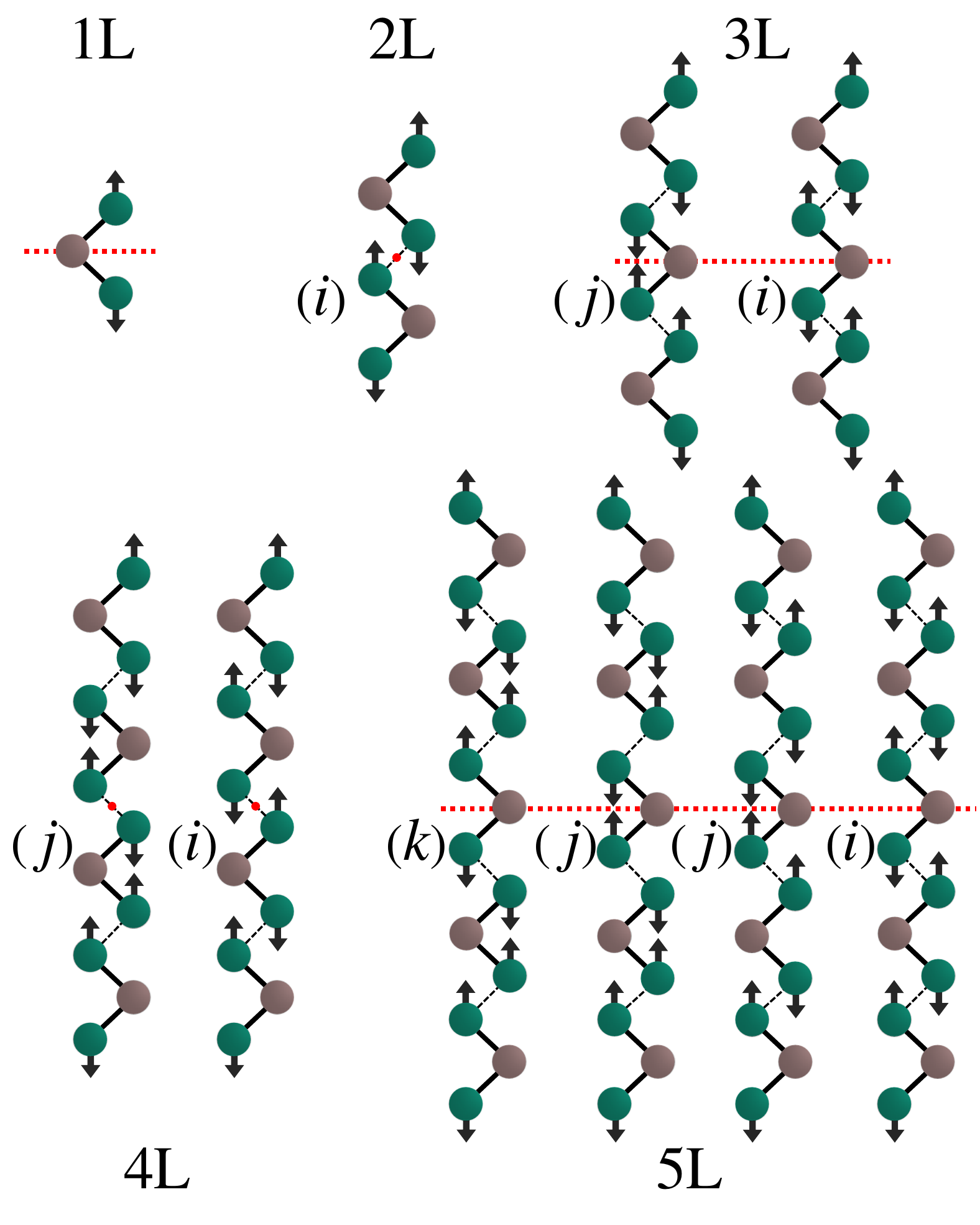}

\caption{(color-online) Schematic representation of out-of-plane Raman active vibrational modes in MoTe$_{2}$ thin layers. The red points mark the inversion centers and the red dashed lines mark the mirror symmetry plane.}

\label{fig:A1g}
\end{figure}

In the case of monolayer MoTe$_{2}$, the single-molecule composition of the unit cell implies the existence of only one mode of the out-of-plane vibrations. 
This mode (A$^{'}_{1}$) is Raman active and as such gives rise to the appearance of an individual Lorentz type line in the Raman scattering spectrum, displayed in Fig. \ref{fig:porownanie}.

Two out-of-plane modes are expected in bilayer MoTe$_{2}$, as there are two molecules in the unit cell (Fig. \ref{fig:A1g}). The modes are similar to those of bulk, except for the number of interacting layers. The mode with Te atoms vibrating in-phase in both planes (A$_{1\text{g}}$) is Raman-active. The other mode, with the 180$^{\text{o}}$ phase difference between the oscillations of Te atoms in two planes (A$_{2\text{u}}$) is IR-active\cite{froehlicher}. 
As a result only one peak should be observed in the Raman scattering spectrum (see Fig. \ref{fig:porownanie}).

There are two Raman active modes in trilayer MoTe$_{2}$ (Fig. \ref{fig:A1g}). Characteristic for the first one are in-phase oscillations of Te atoms in all three layers of the crystal. The other Raman active mode is due to vibrations in the central layer, which are shifted 180$^{\text{o}}$ with respect to the vibrations in the outer layers. For the sake of the present work we label the mode with all Te atoms vibrating in-phase with ($i$). The mode in which the nearest neighbour Te atoms in adjacent layers vibrate out-of-phase will be denoted with ($j$).
For trilayer MoTe$_{2}$, the energies of these modes are expected to be arranged almost symmetrically around the energy of the monolayer (A$^{'}_{1}$) mode, which in fact is observed in the experiment. 

Two out-of-plane modes, which have an inversion symmetry are expected to be Raman active in tetralayer MoTe$_{2}$. A first mode with the Te atoms in all four layers vibrating in-phase - ($i$) and a second one - ($j$), in which the oscillations of the Te atoms in the two outer planes are shifted by 180$^{\text{o}}$ with respect to their vibrations in the two central layers (Fig. \ref{fig:A1g}). This results in two Raman-active modes as observed in the spectrum (see Fig. \ref{fig:porownanie}).

There are four Raman active modes in pentalayer MoTe$_{2}$. One of them, ($i$), corresponds to Te atoms vibrating in-phase in all layers of the crystal. In the other three modes the Te atoms in one, two or three inner layers oscillate out-of-phase as compared to the Te atoms in the outer planes (Fig. \ref{fig:A1g}). Identical interactions involved in the modes with one and three layers vibrating out-of-phase lead to their degeneracy. Those modes will be collectively denoted with ($j$). There are two inter-layer spaces in which Te atoms in adjacent layers vibrate out-of-phase in the mode ($j$).
On the contrary in the mode with two inner layers oscillating out-of-phase as compared to the Te atoms in the outer planes, there are four interlayer-spaces in which the nearest neighbour atoms in adjacent planes move out-of-phae to each other. The corresponding mode will be labelled with ($k$). As the energy of the mode depends on the number of interlayer spaces in which Te atoms in adjacent layers vibrate out-of-phase (zero, two, and four in the ($i$), ($j$), and ($k$) modes respectively) only three different features should be observed in the Raman scattering spectrum of pentalayer MoTe$_{2}$.
This is borne out by the experiments and seen in Fig. \ref{fig:porownanie}.

The increase in the number of layers results in the increase in the number of possible vibrational modes (refer to Fig. \ref{fig:porownanie}). Keeping in mind that Raman active modes must preserve the mirror symmetry with respect to the plane cutting the Mo atoms in the central layers (for odd $N$) or the inversion symmetry at the inversion center (for even $N$ - as indicated by the red dot in Fig. \ref{fig:mody} for the analogous A$_{1\text{g}}$ mode in bulk) one may predict the number of measurable modes as a function of $N$. The degeneracies of some of these modes can be figured out from the analysis of the interactions between adjacent layers. The mode ($i$), corresponding to the in-phase movement of Te atoms in all of the layers manifests itself as the highest energy peak from the group of lines gathered around 170 cm$^{-1}$ in MoTe$_{2}$ flakes of any given thickness. The modes ($j$), ($k$), ($l$) ... correspond to Te atoms in adjacent layers vibrating out-of-phase  in two, four, six, .... inter-layer spaces.

\begin{figure}[h]
\includegraphics[width=.48\textwidth]{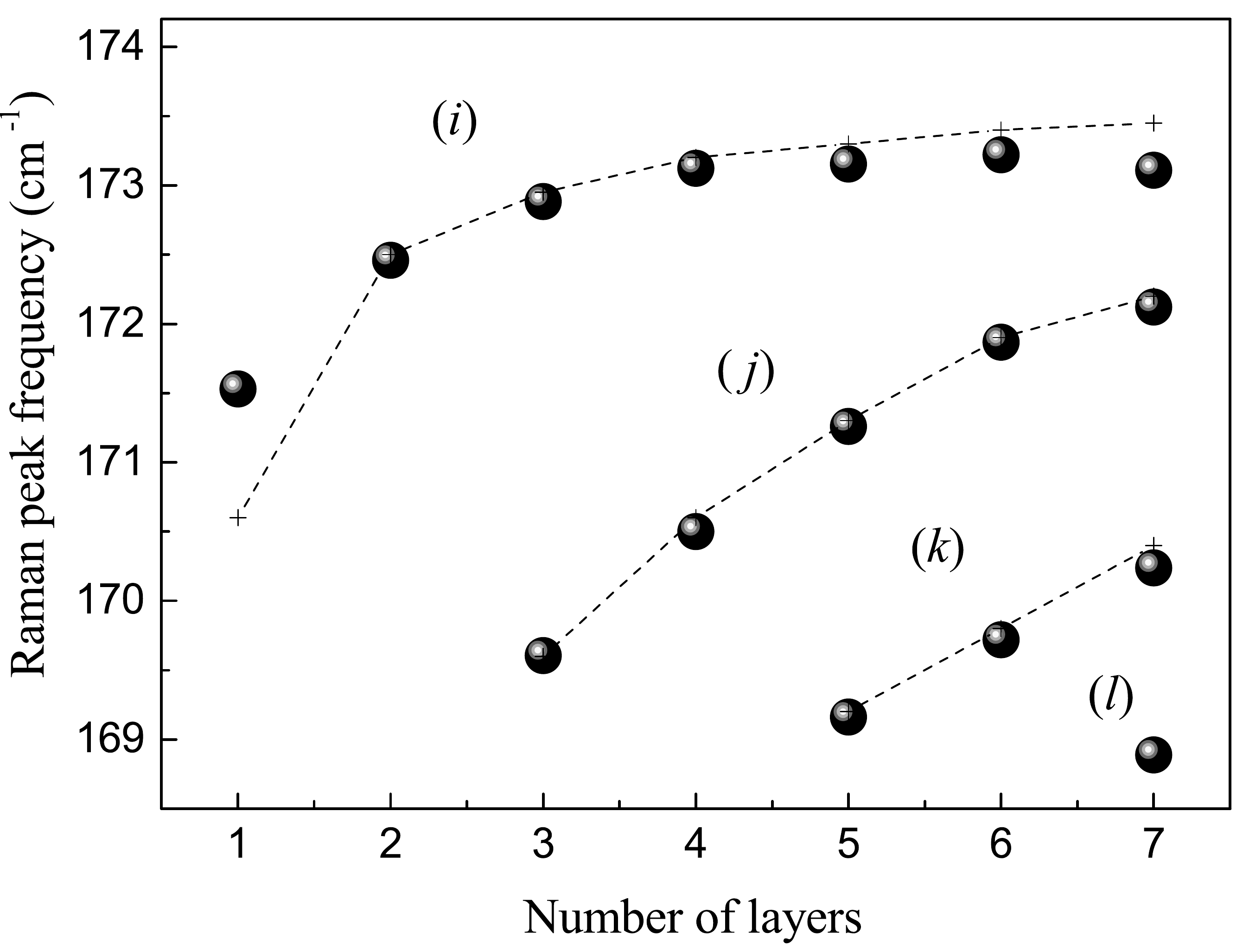}

\caption{The frequency evolution of the Raman peaks related to the scattering by out-of-plane vibrations at room temperature in mono- up to septa-layer MoTe$_{2}$ excited with 632.8~nm laser light. 
The closed circles represent our experimental data. 
The cross symbols connected with dashed lines correspond to model presented in the work of G. Froehlicher et al.\cite{froehlicher}.}

\label{fig:ewolucja}
\end{figure}

The energies of the spectral lines in the Raman scattering spectra of mono- to septa-layer MoTe$_{2}$ excited with 1.96 eV energy, which are related to out-of-plane vibrational modes are presented in Fig. \ref{fig:ewolucja}. As it can be seen in the Figure \ref{fig:ewolucja} their energy evolution with the number of layers agrees with the model presented in the work of G. Froehlicher et al.\cite{froehlicher}

\begin{figure}[h]
\includegraphics[width=.5\textwidth]{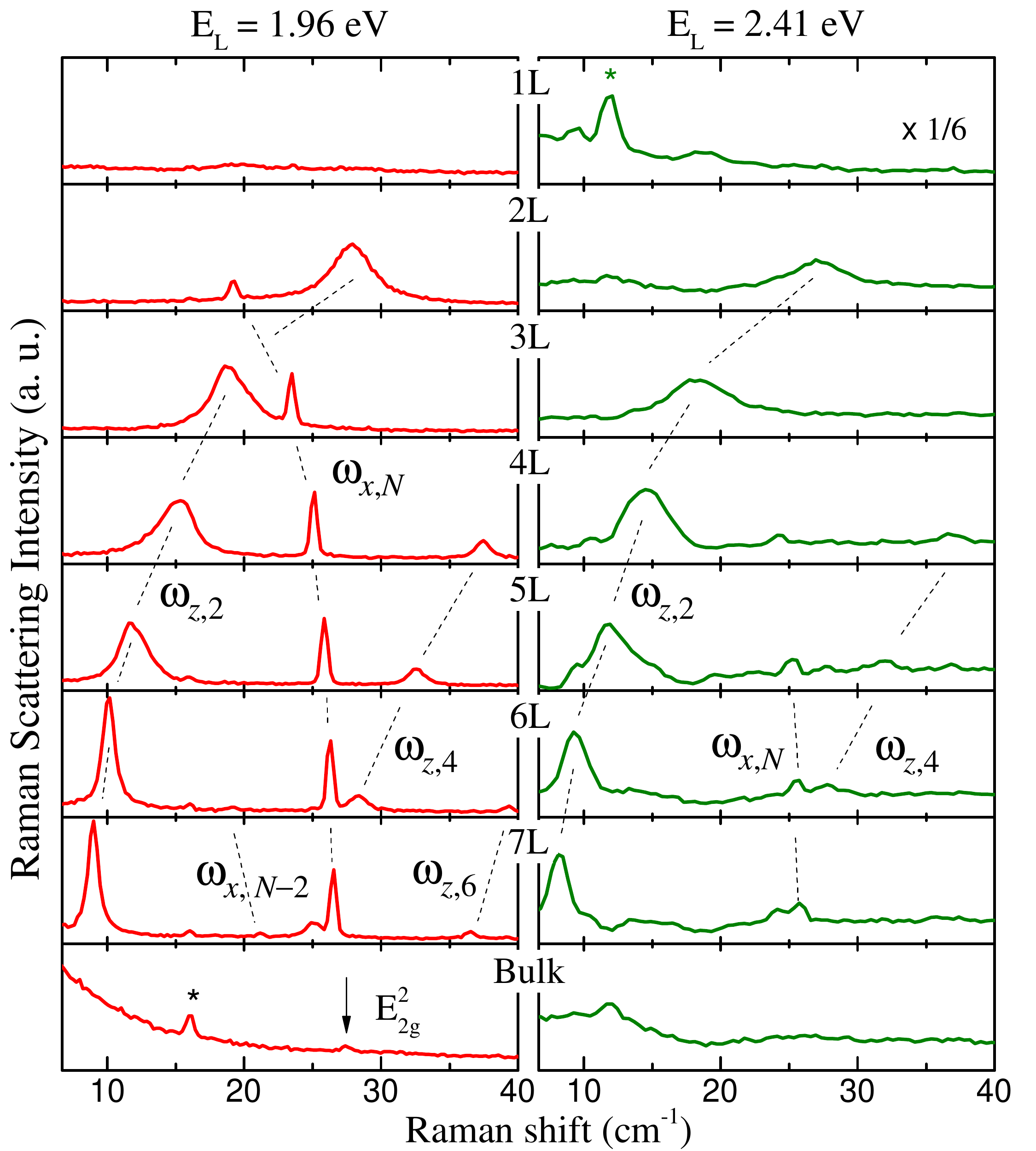}

\caption{(color online) Low-energy Raman scattering spectra of mono-layer (1L) up to septa-layer (7L) and bulk MoTe$_{2}$ measured at room temperature with the use of 632.8 nm and 514.5 nm laser light excitation. 
The feature denoted with an asterisk in the 1L spectrum excited with 514.5 nm light is related to laser, for clarity it was subtracted from the other spectra. The feature denoted with an asterisk in the spectrum excited with 632.8 nm light from the bulk (present also in the spectra for 5L, 6L, and 7L) is related to laser.}
\label{fig:SB}
\end{figure}

\begin{figure}[h]
\includegraphics[width=.48\textwidth]{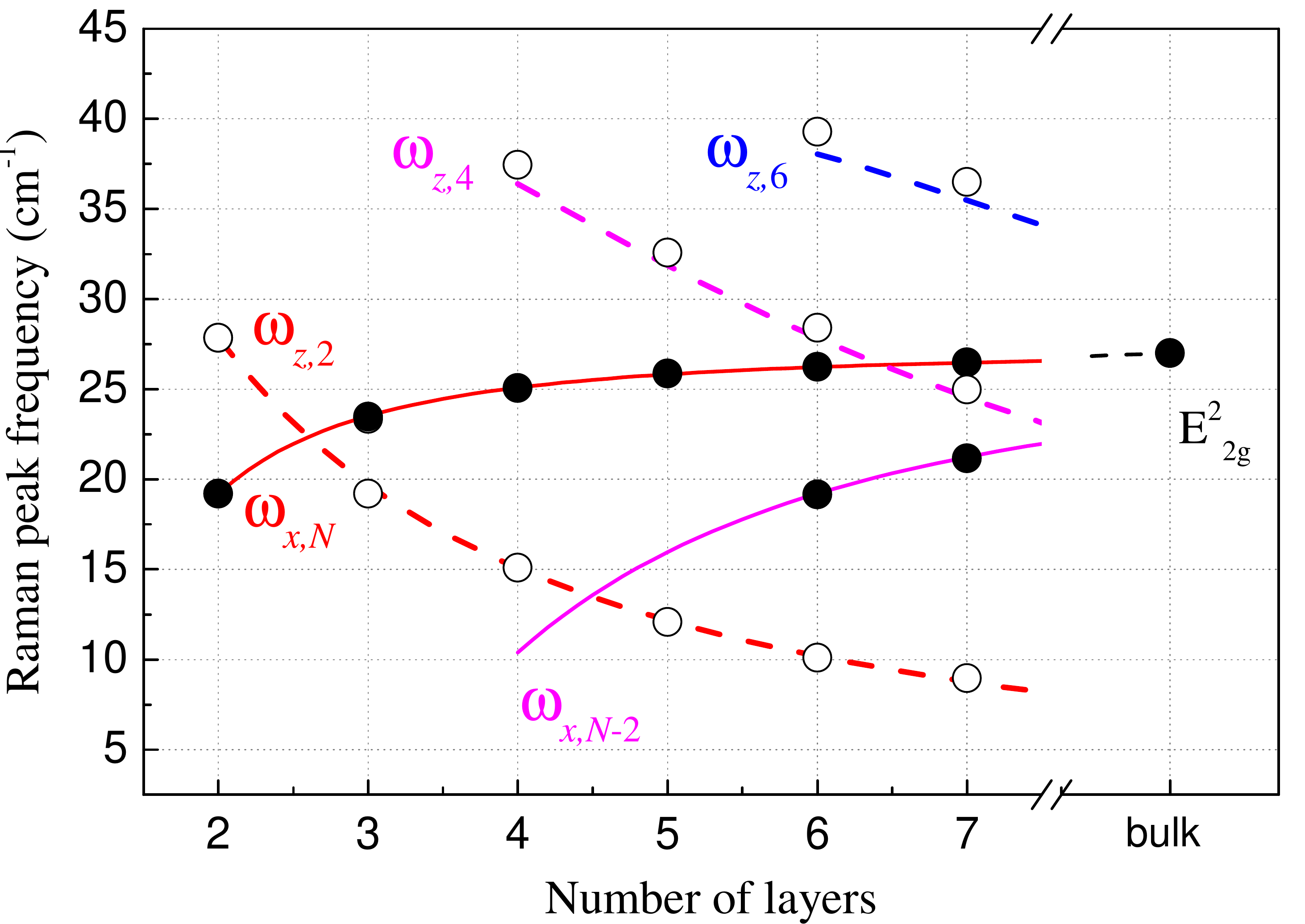}
\caption{(color online) Thickness dependence of the peak positions of the modes observed at room temperature in the low-energy Raman scattering spectra of mono- to septa-layer and bulk MoTe$_{2}$ excited with 632.8 nm laser light. The circles represent the energies extracted from experimental data. The experimental points corresponding to shear (breathing) modes are shown with solid (open) circles.
The theoretical evolution of the shear (breathing) modes is presented with the solid (dashed) curves.}
\label{fig:E22g}
\end{figure}

In order to make our study more complete we also measured the Raman scattering spectra of few-layer and bulk MoTe$_{2}$ in the low energy range (below 60 cm$^{-1}$). 
As it is known from the research on MoTe$_{2}$ \cite{froehlicher} and other layered materials \cite{zhang2,zhao3} one may expect in that energy range two families of modes of interlayer oscillations, i.e. shear and breathing modes, related to rigid layer displacements which are perpendicular and parallel to the $c$-axis, respectively. 
Our experimental results obtained under 632.8~nm and 514.5 nm light excitation are summarized in Fig. \ref{fig:SB}. 
As expected, regardless of the energy of the exciting light there are no rigid interlayer vibrations in the case of monolayer MoTe$_{2}$ and two modes (one shear and one breathing) exist in the bilayer. 
With increasing material thickness the number of features present in the experimental data also increases. 
This effect is especially pronounced for the excitation with 632.8 nm light (see Fig. \ref{fig:SB}). 
Finally in bulk only one peak corresponding to the Raman-active E$^{2}_{2\text{g}}$ shear mode remains in the spectrum. 
The energy positions of all the peaks observed in the low-energy Raman scattering spectra of mono- to septa-layer and bulk are summarized in Fig. \ref{fig:E22g}. 
In order to understand their evolution with the number of layers, we analyze them using the linear chain model of the crystal lattice vibrations. 
This approach has recently been proven to provide a reasonable description of the rigid oscillation modes in MoTe$_{2}$ \cite{froehlicher}, MoS$_{2}$ and WeSe$_{2}$ \cite{zhang2,zhao3}. Every triple (Te--Mo--Te) layer of atoms forming MoTe$_{2}$ crystal is represented in the model by a single point with the reduced mass per unit area $\mu=2m_{\text{Te}}+m_{\text{Mo}}$, where $m_{\text{Te}}=2\cdot10^6 \frac{\text{kg}}{\text{m}^{2}}$ and $m_{\text{Mo}} = 1.5\cdot10^{6} \frac{\text{kg}}{\text{m}^{2}}$ stand for the masses per unit area of the tellurium (Te) and molybdenum (Mo) atoms, respectively. 
Only the interaction between adjacent layers is taken into account in the calculations, which is described by a single force constant $K_{i}$ equal to the in-plane force constant per unit area $K_{x}$ in the case of shear modes and to the out-of plane compression force constant per unit area $K_{z}$ for the breathing modes.

The evolution of the interlayer mode energies (in cm$^{-1}$) as a function of $N$ is represented within the model by the following equation:

\begin{align*}
\omega_{i,\alpha} & = \sqrt{\frac{K_{i}}{2\mu\pi^{2}c^{2}}\left ( 1-\cos\frac{(\alpha-1)\pi}{N}\right )} = \\
 & = \omega_{i} \sqrt{1-\cos\frac{(\alpha-1)\pi}{N}}
\end{align*}

$\text{where } \alpha=2, 3, \ldots, N$

\qquad

As can be noticed there is one set of phonon branches expected for the shear and one for the breathing modes. 
The Raman active shear (breathing) modes belong to the higher-energy (lower-energy) branches of the corresponding set. 

We start our analysis with the shear mode which can be observed in 2L MoTe$_{2}$ at 19.2 cm$^{-1}$. 
Assuming that the force constant $K_{x}$ does not significantly change as $N$ increases from 2 to infinity \cite{zhao3} we computed the thickness dependence of the shear modes with $\alpha = N, N-2$. 
As can be seen in Fig. \ref{fig:E22g}, the model reasonably reproduces the observed evolution of the modes, which stiffen with increasing number of layers. 
The $\omega_{x,N}$ and  $\omega_{x,N-2}$  branches correspond to previously observed S$_{1}$ and S$_{2}$ shear modes in MoS$_{2}$ and WSe$_{2}$ \cite{zhao3}, which are also referred to as $\text{C}^{+}_{2}$ and $\text{C}^{+}_{6}$. \cite{zhang}

Knowing the value of $\omega_{x}$ (19.2 cm$^{-1}$) one can find the in plane (shear) force constant $K_{x}$, which in our case is equal to $3.6\cdot10^{19} \frac{\text{N}}{\text{m}^{3}}$ which is somehow larger than the values for WSe$_{2}$ ($3.1\cdot10^{19} \frac{\text{N}}{\text{m}^{3}}$) and MoS$_{2}$ ($2.7\cdot10^{19} \frac{\text{N}}{\text{m}^{3}}$). \cite{zhao3} 
The force constant $K_{x}$ can also be used to determine the corresponding elastic constant C$_{44}$. Assuming the distance between the centers of neighboring MoTe$_{2}$ layers $t = 0.72$ nm, one can calculate  C$_{44}=K_{x}\cdot t = 25$~GPa. 

\noindent{The results are also in agreement with the recently reported values for MoTe$_{2}$.\cite{froehlicher}}

A similar analysis of the breathing modes starts with the energy $\omega_{z} = 27.8$ cm$^{-1}$, the energy of the breathing mode observed in bilayer MoTe$_{2}$ (see Fig. \ref{fig:SB}). The breathing modes soften (redshift) with increasing number of layers. The theoretically predicted thickness evolution of the $\omega_{z,2}$, $\omega_{z,4}$, and $\omega_{z,6}$ breathing-mode subbranches is shown in Fig. \ref{fig:E22g} with dashed lines. It can be seen that the simulated curves fit fairly well the observed low-energy peaks. The resulting out-of-plane (breathing) force constant $K_{z}$, which equals $7.5 \cdot 10^{19} \frac{\text{N}}{\text{m}^{3}}$ is in reasonable agreement with the recently published data for MoTe$_{2}$ ($7.8 \cdot 10^{19} \frac{\text{N}}{\text{m}^{3}}$) \cite{froehlicher} and it is smaller than the value for WSe$_{2}$ or MoS$_{2}$ ($8.6 \cdot 10^{19} \frac{\text{N}}{\text{m}^{3}}$). \cite{zhao3} The corresponding elastic constant C$_{33}$ in MoTe$_{2}$ can be determined as $K_{z}\cdot t = 54$~GPa. 

The most prominent feature of the results presented so far is the rich structure of the Raman spectra of MoTe$_{2}$ both for the peaks related to the out-of-plane vibrations at the $\Gamma$  point of the Brillouin zone and for the low-energy vibrations. Additionally, there is also a substantial difference between the spectra excited with 1.96 eV and 2.41~eV light energy (see Figs \ref{fig:porownanie} and \ref{fig:SB}). For 5L and 6L (7L) three (four) components of the out-of-plane vibrations can be seen in the spectrum excited with the 632.8~nm light, while just one peak can be observed in the 514.5~nm light-excited spectrum. Moreover, with the 514.5 nm excitation the highest energy component of the spectrum due to out-of-plane vibrations $(i)$, which corresponds to the in-phase vibrations in all layers, has the highest intensity. 
This is not the case of the 632.8 nm excited spectrum, in which, for $N>2$, the ($i$) mode leads to the weakest feature in the spectrum. Similarly the shear modes in the low-energy spectrum excited with 1.96 eV light energy are more clearly seen that in the higher energy (2.41~eV) excited spectrum. In our opinion the difference results from the resonant character of the Raman scattering excited with 632.8 nm light. In order to confirm the effect, we compared the intensities of the Raman scattering for Stokes and anti-Stokes modes. It is well known that assuming a Bose-Einstein occupancy of the phonon states, the ratio of anti-Stokes ($I_{\text{AS}}$) and the Stokes ($I_{\text{S}}$) intensities of the spectral line of the frequency $\omega_{ph}$ can be expressed as:

\begin{equation} \label{eq:1}
\frac{I_{AS}}{I_{S}} = \text{exp} \left ( - \frac{\hslash \omega_{ph}}{kT^{*}} \right )
\end{equation}

\begin{figure}[h]
\includegraphics[width=.49\textwidth]{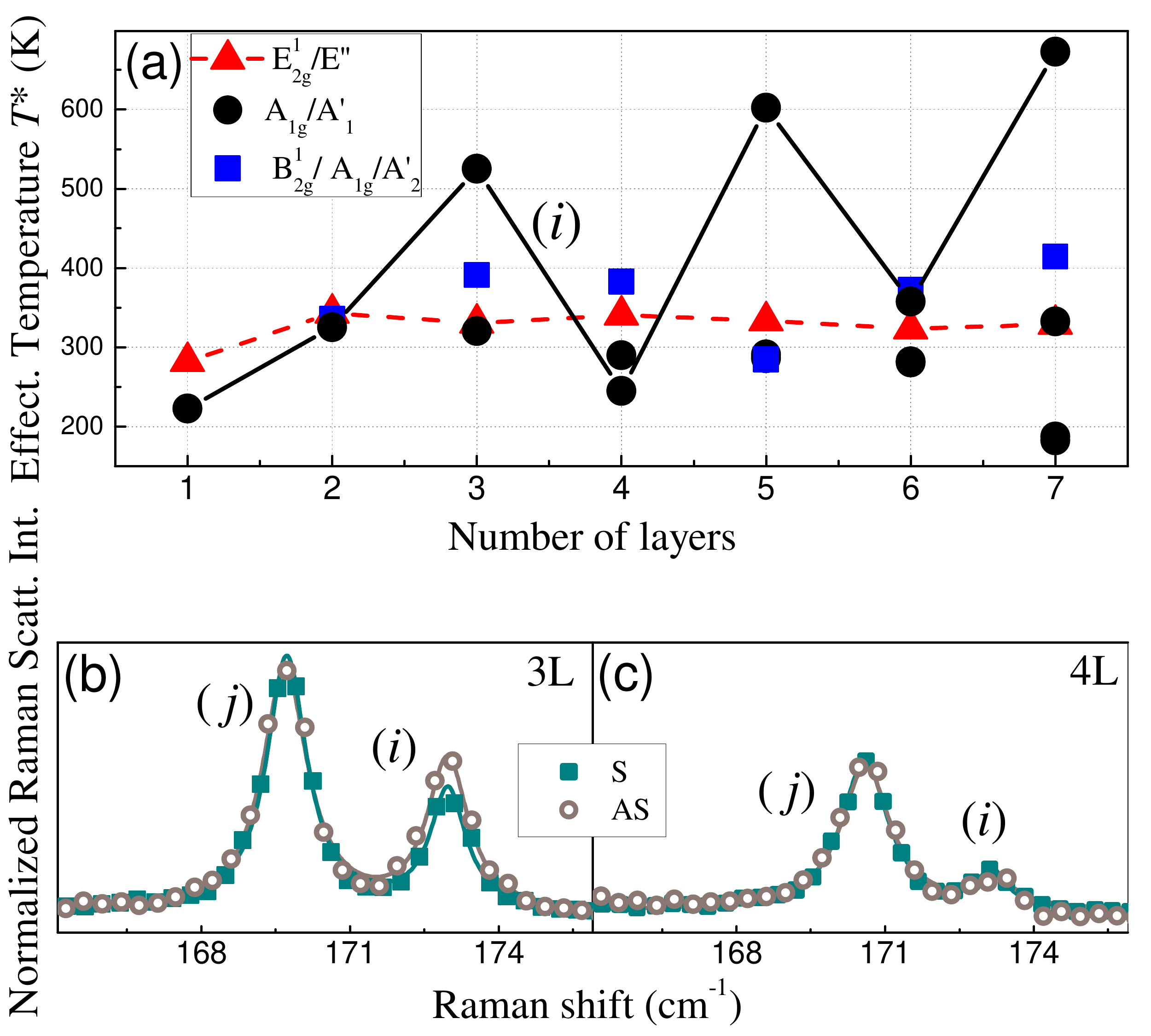}

\caption{(color online)(a) Effective temperature $T$* determined for the Raman peaks in mono- to septa-layer measured at room temperature with the use of 632.8 nm light excitation. 
The points, which correspond to the ($i$) modes are connected with a continous line for more visibility. 
The anti-Stokes (AS) and Stokes (S) spectra due to the out-of-plane vibrations in trilayer (b) and tetralayer (c) are also shown as circles (AS) and squares (S). 
The spectra are normalized at the ($j$) mode to highlight the enhancement of the ($i$) mode in the AS spectrum of 3L MoTe$_{2}$.
The solid curves represent the results of fitting the experimental data with a superposition of Lorentzian curves.}

\label{fig:temperatura}
\end{figure}

The effective temperature $T^{*}$ determined from Eq. \ref{eq:1} for all the observed features is summarized in Fig. \ref{fig:temperatura}(a). It can be seen that for E$^{1}_{2\text{g}}$/E$^{'}_{2}$ modes as well as for the modes corresponding to B$^{1}_{2\text{g}}$ in bulk MoTe$_{2}$ the effective temperature $T^{*}$ stays in reasonable agreement with room temperature of the experiment. In our opinion this confirms that the possible heating effect of the laser light is not significant in our experiment. On the contrary there are substantial discrepancies between the effective temperatures $T$* determined from Eq. \ref{eq:1} for the ($i$) modes in 3L, 5L, and 7L. The effect can be appreciated in Figs \ref{fig:temperatura}(b) and (c) in which the anti-Stokes and Stokes spectra for 3L and 4L are compared. The spectra are normalized at the maximum of the ($j$) modes in the corresponding spectra. It can be clearly seen in Fig. \ref{fig:temperatura}(b) that the relative intensity of the anti-Stokes ($i$) mode in 3L MoTe$_{2}$ is larger than the relative intensity of the ($j$) mode.
This corresponds to the effective temperature determined for the mode ($i$) which is higher than temperature predicted form the Bose-Einstein distribution.


In our opinion the discrepancy between the effective temperature and the equilibrium temperature of the crystal lattice for the ($i$) modes suggests the resonant effect of the 1.96 eV light excitation on the out-of-plane vibrations in thin MoTe$_{2}$ layers. Two observations should be addressed in order to explain the effect. First is the fact that only the ($i$) mode is enhanced in the anti-Stokes spectrum. In our opinion the ($i$) peak couples with the electronic resonance while the other mode, ($j$), does not. The energy of the exciting photons (1.96 eV) coincides with the maximum of the electronic density of states (2.07 eV in the bulk) in the highest valence band and the second lowest conduction band at the M point of the Brillouin zone\cite{guo}. As it can be appreciated in Ref.~\citenum{ruppert}, the edge states at the M point in MoS$_{2}$ have substantial component of S $p_{z}$ orbitals, which point along $c$-axis of the crystal. A similar band-structure, with the Te $p_{z}$ orbitals contributing to the maximum of the valence band in M point of the Brilloiun zone can be expected in MoTe$_{2}$. In the ($i$) mode the distance between Te atoms in adjacent layers modulates. The resulting modulation of the interaction between Te-related $p_{z}$ orbitals affects the MoTe$_{2}$ band-structure near the M-point of the Brilloiun zone. In our opinion this results in the strong electron-phonon coupling between the electronic transition at M point of the Brilloiun zone and the ($i$) mode of the out-of-plane vibrations. In the ($j$) mode Te atoms in adjacent layer vibrate out-of-phase and the modulation of their relative distance is weaker, which may explain the weaker electron-phonon coupling. Another point to address is related to the different effect of 1.96 eV light on the ($i$) modes in odd- and even-$N$ MoTe$_{2}$. This may be related to a specific effect of the crystal symmetry in odd - $N$ layer TMDs (lacking the inversion center) on their energy structure. In our analysis of the effect of 1.96 eV light excitation on the spectrum of the out-of-plane vibrations we also note that the excitation energy coincides with the energy of B' exciton in MoTe$_2$.\cite{ruppert}
One could expect an effect similar to the resonant enhancement of the A$_{1\text{g}}$ mode in MoS$_2$ with the excitation coinciding with the energy of A and B excitons.\cite{livneh, carvalho}
This however would not clearly explain the different effect of the excitation on different modes of the out-of-plane vibrations in MoTe$_{2}$. 

In order to explain our results, more strict theoretical analysis is definitely needed, which is beyond the scope of our experimental study. We belive however that our report would encourage that type of studies for the sake of the expected interesting physics behind. 


\section{Conclusions \label{sec:summary}}

We have carried out a study on room temperature Raman scattering in few-layer flakes of exfoliated MoTe$_{2}$ using 632.8 nm laser light excitation. Our experimental data show that out-of-plane vibrational modes (corresponding to the A$_{1\text{g}}$ mode in the bulk) in such flakes posses a complex structure with the peaks related to the interlayer interactions between the single planes of the MoTe$_{2}$. We have also demonstrated that the energies of the low-energy modes of rigid interlayer vibrations of MoTe$_{2}$ crystal lattice evolve with the number of layers in a way that can be satisfactory reproduced within a linear chain model with only the nearest neighbor interactions taken into account. We argue that the rich structure appears because of the resonance between the incoming photons with the energy gap between the maximum of the valence band and the minimum of the second lowest conduction band at the M point of the MoTe$_{2}$ first Brillouin zone.

After the submission of the present manuscript the authors became aware of a report with similar experimental results by Q.J.Song et al, published in Ref. \citenum{song2016}.

\begin{acknowledgments}

This work has been supported by the National Science Center under grants no. DEC-2013/11/N/ST3/04067, DEC-2013/10/M/ST3/00791 and DEC-2015/16/T/ST3/00496. Funding from European Graphene Flagship and European Research Council (ERC-2012-AdG-320590-MOMB) is also acknowledged. We kindly acknowledge the support from the Nanofab facility of the Institut Néel, CNRS UGA.

\end{acknowledgments}

\bibliographystyle{apsrev4-1}
\bibliography{quabi}

\end{document}